\documentclass[aps,preprint]{revtex4}
\usepackage{ifpdf}
\ifpdf
\usepackage[pdftex]{graphicx}
\usepackage{epstopdf}
\else
\usepackage{graphicx}
\fi
\usepackage[nolist]{acronym}
\usepackage{float}
\usepackage{calc}
\usepackage[nomessages]{fp}
\usepackage{subfig}
\restylefloat{table}

\begin{acronym}
\acro{Beem}[BEEM]{Ballistic electron emission microscopy}
\acro{beem}[BEEM]{ballistic electron emission microscopy}
\acro{XPS}{x-ray photo-emission spectroscopy}
\acro{STM}{scanning tunneling microscopy}
\acro{AFM}{atomic force microscopy}
\acro{AES}{auger electron spectroscopy}
\acro{RBS}{Rutherford backscattering spectrometry}
\acro{SBH}{Schottky barrier height}
\acro{FLT}{Fermi liquid theory}
\acro{UHV}{ultra high vacuum}
\acro{IBS}{interface band structure}
\acro{MBE}{molecular beam epitaxy}
\acro{BK}{Bell and Kaiser}
\acro{PL}{Prietsch Ludeke}
\end{acronym}

\begin{document}
\title{Relating Spatially Resolved Maps of the Schottky Barrier Height to
Metal/Semiconductor Interface Composition}
\author{Robert Balsano}
\author{Chris Durcan}
\author{Akitomo Matsubayashi}
\affiliation{College of Nanoscale Science and Engineering,
University at Albany, SUNY, Albany, New York 12203, USA}

\author{Vincent~P.~LaBella}\email{vlabella@albany.edu}
\altaffiliation{Present Address: SUNY Polytechnic Institute, Albany, New York
12203, USA}
\affiliation{College of Nanoscale Science and Engineering,
University at Albany, SUNY, Albany, New York 12203, USA}

\date{\today}

\begin{abstract}

The \ac{SBH} is mapped with nanoscale resolution at pure Au/Si(001) and mixed
Au/Ag/Si(001) interfaces utilizing \ac{beem} by acquiring and fitting spectra
every 11.7~nm over a $1~\mu$m$~\times~1~\mu$m area. The energetic distribution
of the \ac{SBH} for the mixed interfaces contain several local maximums
indicative of a mixture of metal species at the interface.  To estimate the
composition at the interface, the distributions are fit to multiple Gaussians
that account for the species, ``pinch-off'' effects, and defects.  This
electrostatic composition is compared to \ac{RBS} and \ac{XPS} measurements to
relate it to the physical composition at the interface.
\end{abstract}



\maketitle
\acresetall
\section{Introduction}

Metal/semiconductor interfaces form rectifying Schottky contacts, widely
utilized in power applications due to their low turn on voltages and high
switching speeds. Schottky diodes are also found in sensor applications as both
gas and optical sensors due in part to their
simplicity~\cite{nienhaus:apl:74:4046, li:am:22:2743}. In addition, Schottky
source/drain contacts are being utilized in sub 20~nm node transistors to
improve scalability~\cite{nishisaka:jjap:42:2009, coss:jvstb:31:021202,
smith:mj:82:261, haimsona:me:92:134, papadatos:me:92:2012, tsai:jes:128:2207}.
This motivates the need to study and understand the nanoscale fluctuations in
the Schottky barrier height at metal semiconductor
interfaces~\cite{tung:mser:35:1, tung:prb:45:509, tung:jvstb:11:1546,
tung:prl:84:6078}.

The \ac{SBH} is the energy offset of the conduction band minimum in the
semiconductor with respect to the metal's Fermi level resulting from the bare
space charge that exists at the interface and within the semiconductor.  The
barrier height is dependent upon the type of metal and semiconductor as well as
interface states and bonding, which can be altered by the fabrication process
such as surface pre-treatments, or unintentional foreign
species~\cite{detavernier:jap:84:3226, vanalme:semst:14:871}. These effects can
vary locally and can cause inhomogeneities in the barrier
height~\cite{olbrich:jap:83:358, olbrich:apl:70:2559,
goh:Nanotechnology:19:445718, sirringhaus:prb:53:15944, palm:prl:71:2224,
durcan:jap:116:023705, palm:apa:56:1}. Electrostatic models that explain
\ac{SBH} inhomogeneity predict that a region of metal with a lower barrier
height will have its barrier height increased when it is surrounded by a metal
with a higher barrier height for a given semiconductor
substrate~\cite{tung:mser:35:1, tung:prb:45:509, tung:jvstb:11:1546,
tung:prl:84:6078, detavernier:jap:84:3226}.  This is dubbed \emph{pinch-off} and
is a function of the size of the surrounded region of metal in relation to the
depletion width of the diode. This is in part a consequence of the continuity in
the electrostatic potential at the interface as it will vary smoothly across
regions with different metals that exhibit different barrier
heights~\cite{tung:mser:35:1, tung:prb:45:509, tung:jvstb:11:1546,
tung:prl:84:6078}.

\begin{figure}
\includegraphics[scale=.9]{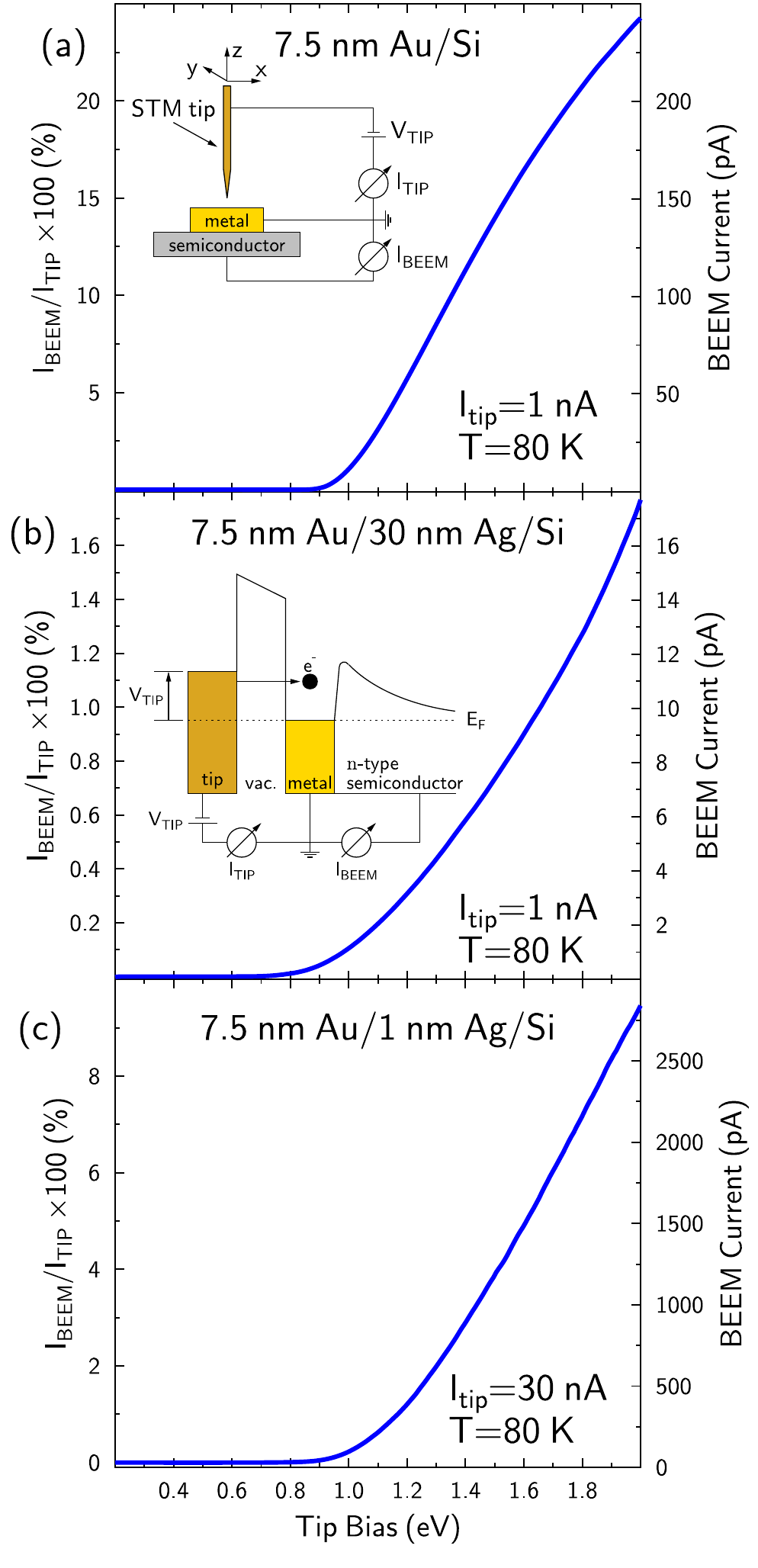}
\caption{\label{all_raw_spectra} Inset (a), the Schematic of the \ac{beem}
experiment, inset (b) band energy diagram of the \ac{beem} experiment, the
averaged \ac{beem} spectrum of (a) 7.5~nm Au/Si(001) sample, (b) 7.5~nm Au/30~nm
Ag/Si(001) sample, (c) 7.5~nm Au/1~nm Ag/Si(001) sample.}

\end{figure}

Fluctuations in the barrier height at the nanoscale can only be examined
utilizing \ac{beem}, which is a powerful technique to measure local \aclp{SBH}
with nanoscale resolution~\cite{bell:prl:61:2368, kaiser:prl:60:1406,
hecht:prb:42:7663}.  It is a three terminal \ac{STM} technique where the
\ac{STM} tip is used to locally inject tunneling electrons into a grounded metal
film deposited onto the surface of a semiconductor as shown schematically inset
in Fig.~\ref{all_raw_spectra}(a)~\cite{bell:prl:61:2368, kaiser:prl:60:1406,
hecht:prb:42:7663}.   Electrons with energy greater than the \ac{SBH} are
collected at the semiconductor and  measured as \ac{beem} current as depicted
inset in Fig.~\ref{all_raw_spectra}(b).  Mapping of the \ac{SBH} is achieved by
fitting the spatially resolved transmission as a function of tip bias to extract
the local \ac{SBH} at individual tip locations and can achieve nanoscale
resolution~\cite{olbrich:jap:83:358, olbrich:apl:70:2559,
goh:Nanotechnology:19:445718, sirringhaus:prb:53:15944, palm:prl:71:2224,
durcan:jap:116:023705, palm:apa:56:1}.

Several studies have mapped the \ac{SBH} using \ac{beem}. Palm et al. took a
series of 12 \ac{beem} images of varying tip bias over a 60~nm~$\times$~60~nm
area on the Au/Si(100) and Au/Si(111) interfaces and did a pixel by pixel fit to
generate a \ac{SBH} map~\cite{palm:prl:71:2224,palm:apa:56:1}.  Olbrich et al.
created \ac{SBH} maps of a mixed Au/Co/GaAs interface consisting of small Co
grains enveloped by Au on a 114~nm~$\times$~114~nm area sampling every 0.89~nm
where pinch off effects were observed~\cite{olbrich:jap:83:358,
olbrich:apl:70:2559}.   Goh et al. mapped \acp{SBH} of an Au/pentacene/Si(111)
interface using a 30~$\times$~30 point grid of spectra taken every
17~nm~\cite{goh:Nanotechnology:19:445718}. Durcan et al. mapped the W/Si(001)
interface with $p$-type and $n$-type substrates over a
1~$\mu$m~$\times$~1$~\mu$m area sampling every
11.7~nm~\cite{durcan:jap:116:023705}.  However, no studies have measured the
Schottky interface of mixed Au and Ag on the Si(001) substrate, which would be
insightful for \emph{pinch-off} effects due to the large differences in their
barrier height's ($\sim$ 0.2~eV) and their
miscibility~\cite{balsano:aipa:3:112110}.

In this article, the \ac{SBH} for Ag and Au mixed films on Si(001) are mapped
with nanoscale resolution by acquiring 7,225 spectra in a regularly spaced grid
over a 1~$\mu$m~$\times$~1~$\mu$m area.  False color spatial maps and energy
histograms of the \ac{SBH} are utilized to relate the electrostatic character of
the interface to its material composition. The distributions are fit to a sum of
Gaussian distributions arising from the presence of multiple species at the
interface, ``pinch-off'' effects, and interface defects.  The composition
obtained from this fitting is corroborated with depth resolved chemically
sensitive techniques.

\section{Experimental}

The Schottky diodes were fabricated under \ac{UHV} using $n$-type single crystal
Si(001) wafers with a resistivity of  100 $\Omega$-cm (phosphorus doped). The
native oxide layer was removed utilizing a standard chemical hydrofluoric acid
treatment immediately prior to loading into a \ac{UHV} ($10^{-10}$~mbar)
chamber~\cite{garramone:apl:96:062105, garramone:jvsta:28:643}. The metal films
were  deposited onto the silicon surface using standard Knudsen cells through a
2~mm~$\times$~1~mm  shadow mask.  Three samples were fabricated with  0~nm,
1~nm, and 30~nm thick silver layers while the gold capping layer was kept at 7.5
nm thick for all samples. Each diode was mounted onto a custom designed sample
holder for \ac{beem} measurements. The holder allowed for the metal film to be
grounded using a BeCu wire and connection of the silicon substrate to the ex
situ pico-ammeter to measure the \ac{beem} current.  Ohmic contacts were
established by cold pressing indium into the backside of the silicon substrate.

A modified low temperature \ac{UHV} \ac{STM} (Omicron) was utilized for all
\ac{beem}  measurements with a pressure in the $10^{-11}$ mbar
range~\cite{krause:jvstb:23:1684}.
The samples were inserted into the \ac{UHV} chamber and loaded onto the STM
stage that was cooled to 80~K for all measurements. Two-point current-voltage
measurements were taken \emph{in situ} with \ac{beem} measurement for each
sample at low temperatures  without ambient light using a Keithley 2400 source
measurement unit to verify rectifying behavior. Pt/Ir \ac{STM} tips,
mechanically cut at a steep angle, were utilized for all \ac{beem} measurements.
\ac{beem} spectra were acquired using a constant tunneling current setpoint of
1~nA for the the sample with no silver and 30~nm of silver, while 30~nA was
utilized for the sample with 1~nm of Ag.  All \ac{beem} spectra were acquired
over a tip bias range of $0.20$~eV to $2.00$~eV at 80~K. A spectrum was taken
every 11.7~nm over a 1~$\mu$m~$\times$~1~$\mu$m area of the metal surface,
resulting in 7,225 spectra for each sample.

Each individual spectrum and the average of all spectrum for each sample were
fit to the simplified \ac{BK} model, $I_B \propto (\phi_b - V_t)^n$, where $I_B$
is the \ac{beem} current, $\phi_b$ is the \ac{SBH}, $V_t$ is the tip bias, and
$n = 2$ is the fitting exponent utilizing a linearization
technique~\cite{balsano:aipa:3:112110}.  The fitting returned a \ac{SBH} along
with the $R^2$ value as an indicator of the quality of the fit.  \ac{SBH}
spatial maps and energetic histograms for the individual spectra were  generated
from these fits. \ac{RBS} was performed to measure film thickness and
composition analysis.  An \ac{XPS} sputter depth profile was performed on the
sample with the 1~nm Ag  layer.

\section{Results}

\begin{figure}
\includegraphics[scale=1]{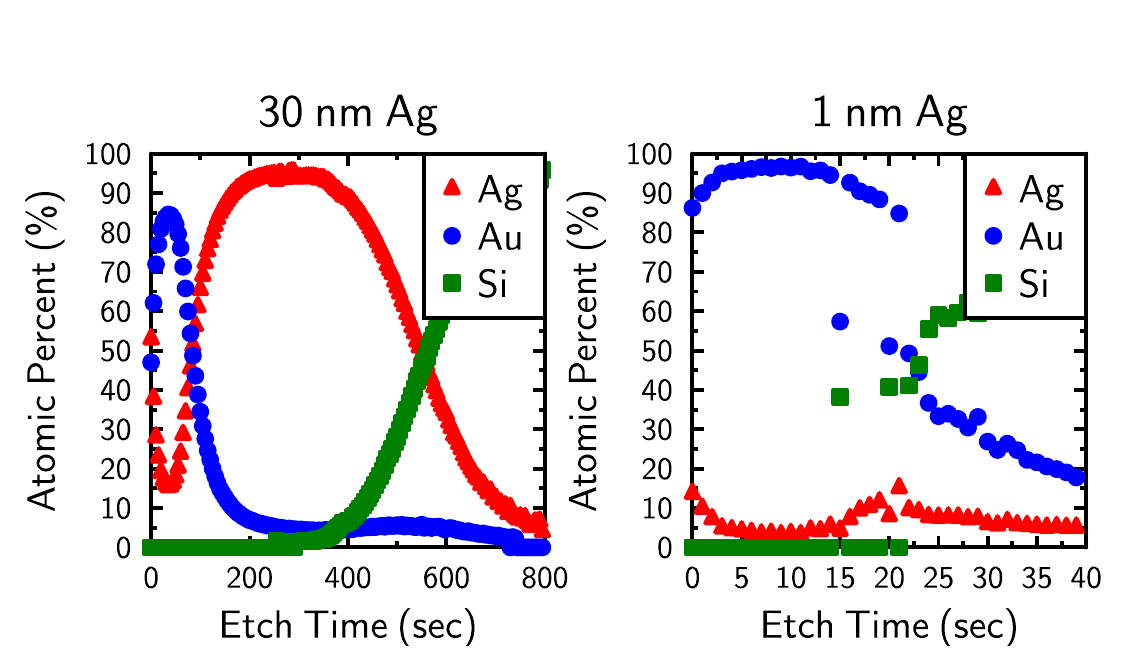}
\caption{\label{depthprofileall}Depth profiles from \ac{XPS} data  for the
7.5~nm Au /30~nm Ag/Si(001)  (left) and 7.5~nm Au /1~nm Ag/Si(001) (right)
diodes.}

\end{figure}

\ac{XPS} sputter depth profiles as a function of time for the  7.5~nm Au/30~nm
Ag/Si  and 7.5~nm Au/1~nm Ag/Si samples are displayed in
Fig.~\ref{depthprofileall} and show the presence of Ag, Au, and Si in the diode.
The composition at the interface is taken to be the composition just before
silicon is detected. In the 30~nm Ag sample, Ag is the dominant species with
95\% at the interface and about 5\% Au.  In the 1~nm Ag sample, Au is the
dominant species with 85\% at the interface and about 15\% Ag.

\begin{figure}
\includegraphics[scale=.7]{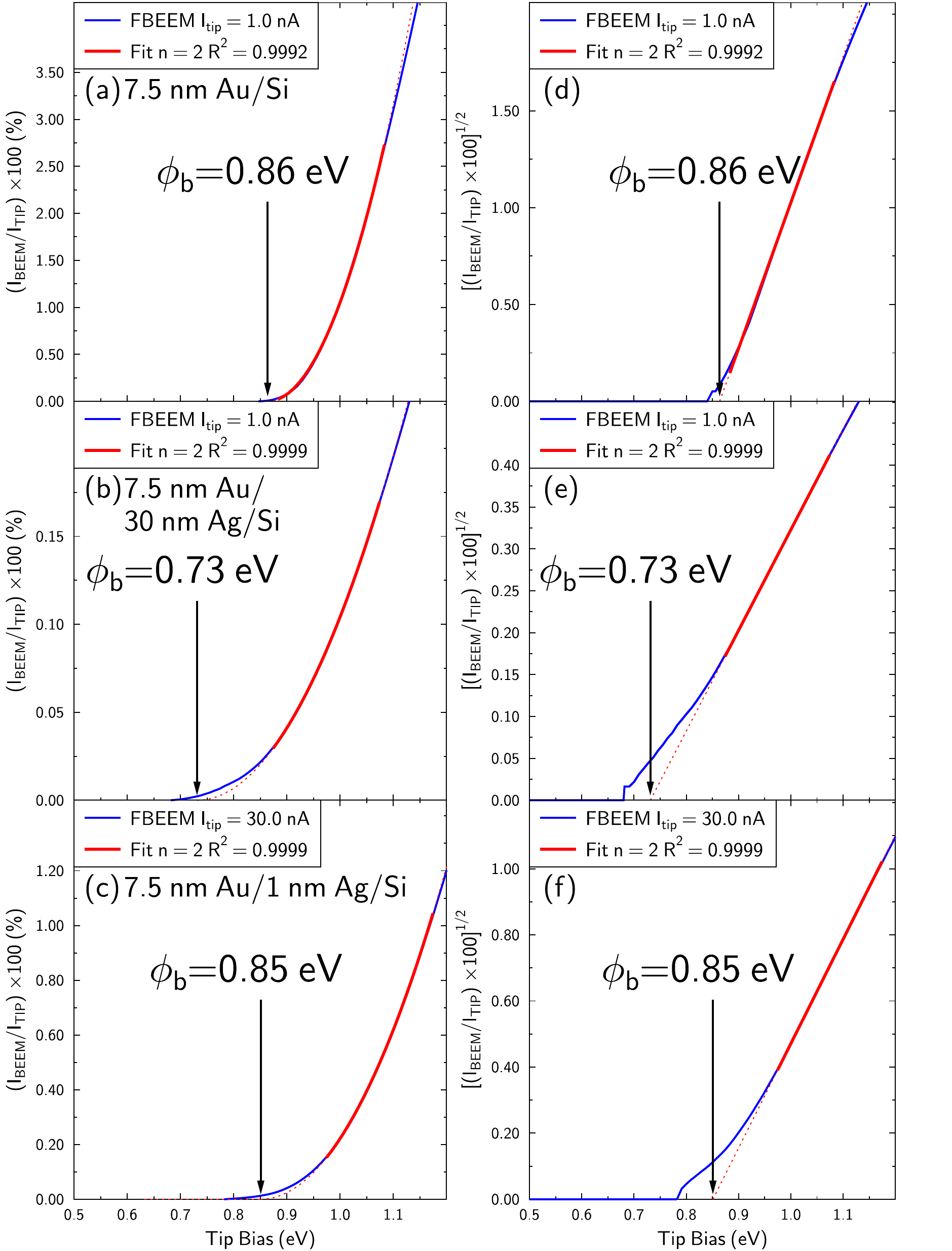}
\caption{\label{allfits} The fits to averaged \ac{beem} spectrum of plotted on a
standard scale (a) 7.5~nm Au/Si(001) sample, (b) 7.5~nm Au/30~nm Ag/Si(001)
sample, (c) 7.5~nm Au/1~nm Ag/Si(001) sample and plotted on as the square root
of their values (d) 7.5~nm Au/Si(001) sample, (e) 7.5~nm Au/30~nm Ag/Si(001)
sample, (f) 7.5~nm Au/1~nm Ag/Si(001) sample.}

\end{figure}

The averaged \ac{beem} spectra for the three samples are displayed in
Fig.~\ref{all_raw_spectra}.  Each spectrum shows an onset threshold
characteristic of a Schottky barrier at the interface. The averaged spectrum and
its fit are shown in raw and linearized forms indicating the \ac{SBH} and R$^2$
value in Fig.~\ref{allfits}. The data is indicated by a blue line, the region of
fit is indicated by a solid red line, and the extrapolation from the region of
fit to the $x$-axis intercept or \ac{SBH} is indicated by a dotted red line.
Visible in the fits is the large difference in the region of extrapolation from
sample to sample where the Au/Ag samples, Fig.~\ref{allfits}~(e) and
Fig.~\ref{allfits}~(f) have extrapolation regions of about 0.14~eV which is
larger than the pure Au sample seen in Fig.~\ref{allfits}~(b).

\begin{figure}
\includegraphics[scale=.25]{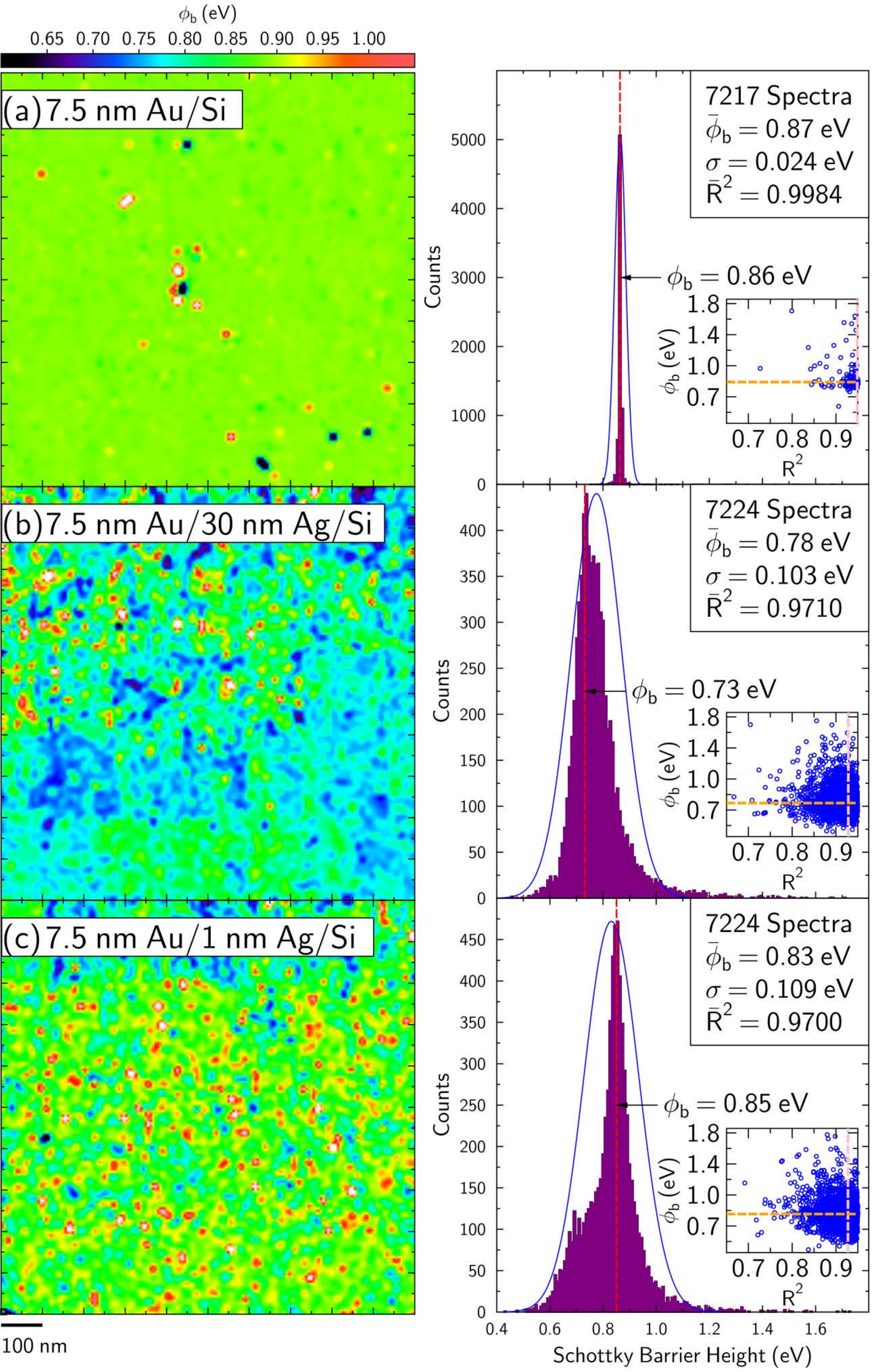}
\caption{\label{allmaps} False color maps of the \acp{SBH} as a function of
position (left) and distribution histograms (right) for (a) 7.5~nm Au/Si(001)
sample, (b) 7.5~nm Au/30~nm Ag/Si(001) sample, (c) 7.5~nm Au/1~nm Ag/Si(001)
sample.  The small box in the bottom left corner of the map in (c) indicates the
pixel size.  The red dotted line on the histograms indicate the \ac{SBH}
calculated from the averaged spectrum.  A normal distribution is drawn around
the data scaled for the size of the bins. Inset in  the histograms, all
\acp{SBH} for the same sample are plotted as a function of $R^2$ value.  The
horizontal yellow dotted line indicates the average \ac{SBH} value.  The
vertical pink dotted line indicates the average $R^2$ value.}

\end{figure}

The \acp{SBH} resulting from fitting each spectrum for all three samples are
displayed as a false color spatial maps (left) and an energetic histogram
(right) in Fig.~\ref{allmaps}~(a)-(c).  A small square is shown in the bottom
left corner to indicate the pixel size of each individual spectrum. The black
pixels indicate a spectrum that could not be fit, while white pixels indicates a
spectrum with a \ac{SBH} above the upper limit of the scale.  The histograms for
all distributions are plotted from 0.40~eV to 1.80~eV, spanning the range of
values obtained from fitting and display the average $R^2$ values of the data
sets. A scatter plot of the \ac{SBH} as a function of $R^2$ value is displayed
in the inset of each histogram. The horizontal and vertical dotted lines mark
the average barrier height and $R^2$ value, respectively. The majority of points
surround the intersect of the two averages and indicate that high barrier
heights have high $R^2$ values.

The spatial distribution of the fitted \acp{SBH} for the 7.5~nm Au/Si sample are
displayed in Fig.~\ref{allmaps}~(a).  The map displays a spatially homogeneous
barrier height near 0.86~eV  with  a few regions where no fit was obtained and
several high barrier regions outside of the color scale.  The histogram
indicates a very narrow distribution of barrier heights extracted from a total
of 7,217 spectra.  The mean barrier height is 0.87~eV with a standard deviation
of 0.024~eV.    The barrier height of the average spectrum, 0.86~eV is indicated
with the vertical red dotted line.

The spatial distribution of \acp{SBH} from the 7.5~nm Au/30~nm Ag/Si diode are
displayed in Fig.~\ref{allmaps}~(b).  The map displays a spatially mixed
distribution of barrier heights with  a small number of locations  where no fit
was obtained and several high barrier regions outside of the color scale.  A
large portion of the image has a barrier height between 0.70~eV and 0.80~eV.
There are also many regions that exhibit a barrier height near 0.86~eV.  The
histogram depicts a broad mixture of barrier heights from 7,224 spectra.  The
mean barrier height is 0.78~eV with a standard deviation of 0.103~eV. The
barrier height from the fit to the average spectrum, 0.73~eV is lower than the
mean and indicated with the vertical red dotted line.

The spatial distribution of \acp{SBH} for the 7.5~nm Au/1~nm Ag/Si sample are
displayed in Fig.~\ref{allmaps}~(c).  The map displays a spatially inhomogeneous
distribution of \acp{SBH} with  a few regions where no fit was obtained and a
number of high barrier regions outside of the color scale.  There are small
patches with \acp{SBH} below 0.80~eV, but the surrounding area is mostly between
0.80~eV and 0.90~eV.  The histogram indicates  7,224 spectra were able to be fit
with the distribution having multiple local maximums.  The mean barrier height
is 0.83~eV with a standard deviation of 0.109~eV.    The barrier height from the
fit to the average spectrum, 0.85~eV is higher than the mean and indicated with
a vertical red dotted line.

Histograms depicting the \ac{beem} data for the 7.5~nm Au/30~nm Ag/Si (a) and
7.5~nm Au/1~nm Ag/Si (b) samples are shown   in
Fig.~\ref{thick_histo_mixed_fit}.    The histogram in (a) illustrates a multi-
modal fit to the distribution in green with its constituent normal distributions
drawn in black.  The means of the five centroids are indicated by red dotted
lines, are $0.62$~eV, $0.73$~eV, $0.76$~eV, $0.86$~eV, and $1.00$~eV, and
contribute  $4.0 \%$, $5.0 \%$, $72.5 \%$, $14.5 \%$, and  $4.0 \%$ to the
multi-modal distribution with standard deviations 0.057~eV, 0.010~eV, 0.053~eV,
0.067~eV, and 0.067~eV, respectively.  The histogram in (b) illustrates a multi-
modal fit to the distribution in green with its constituent normal distributions
drawn in black.  Red dotted lines indicate the distribution means of $0.62$~eV,
$0.69$~eV, $0.75$~eV, $0.85$~eV, and $0.94$~eV, contribute  $3.5 \%$, $7.0 \%$,
$16.0 \%$, $58.0 \%$, and  $15.5 \%$ to the multi-modal distribution and  have
standard deviations 0.041~eV, 0.031~eV, 0.041~eV, 0.036~eV, and 0.072~eV,
respectively.

\section{Discussion}

The \ac{SBH} obtained for the Au sample is in good agreement with the previously
reported values, but the \acp{SBH} for the silver samples are not in agreement
with the reported value of 0.66~eV~\cite{balsano:aipa:3:112110}. The \ac{SBH} of
the average spectrum for the 30~nm Ag sample is over 50~meV higher and almost
200~meV higher for the 1~nm Ag sample. In addition, the 13~meV width of the
extrapolation region for the linearized fits for both Ag samples and the
underestimation of the fit line in the raw spectra indicate that single
threshold fitting is not appropriate.   These effects on the average spectra are
attributed to physical intermixing of Au and Ag at the interface which would
contribute to spectra with different barrier heights being included in the
average spectra.  Evidence of both Au and Ag at the interface for both silver
samples was confirmed with \ac{XPS} and \ac{RBS} depth profiling and is
consistent with the complete miscibility of Ag/Au, which would account for
diffusion of Au to the interface~\cite{wei:prb:36:4163}.

A better view of the effects of this mixing on the electrostatic character of
the interface can be seen in the \ac{SBH} histograms and maps.  The standard
deviation of the histogram of the \ac{SBH} in the Au sample is 24~meV. In
comparison, the \ac{SBH} histograms of both silver samples show highly
asymmetric distributions that are about four times broader.  In addition, the 1
nm silver sample has a secondary peak near 0.7~eV.  The \ac{SBH} spatial maps of
the silver samples show the physical grouping of the high and low barrier
heights, which would be consistent with multiple species at the interface.

\begin{figure}
\includegraphics[scale=1]{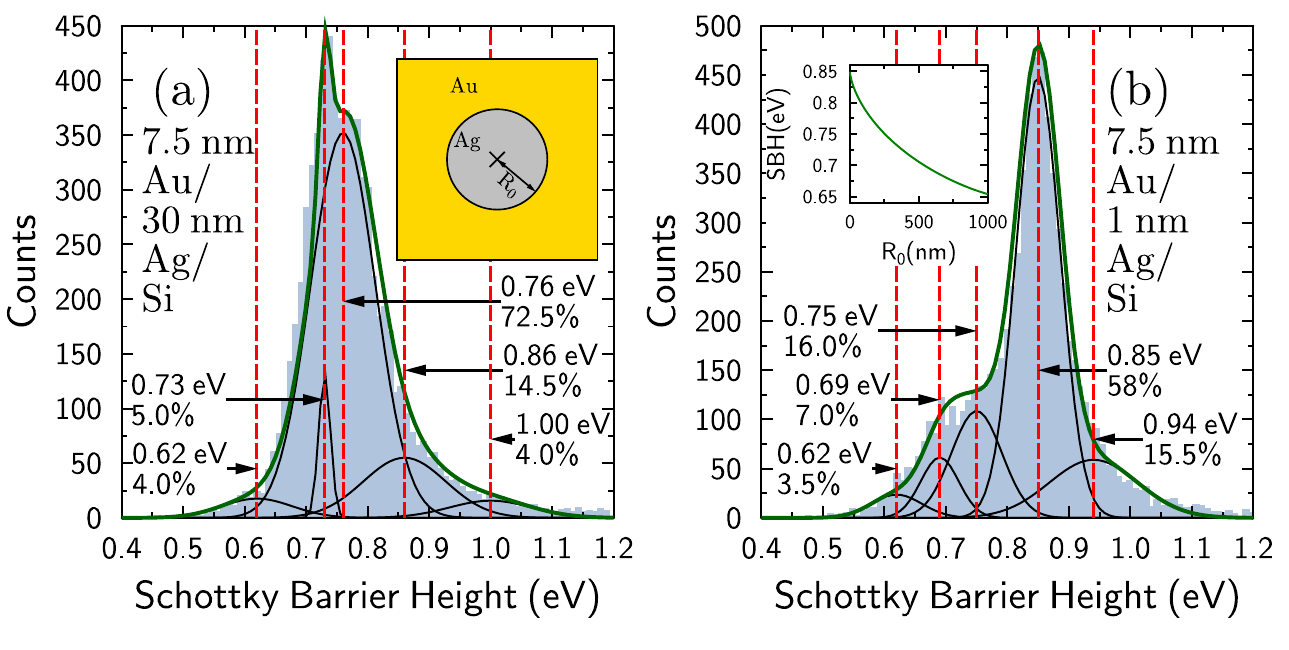}
\caption{\label{thick_histo_mixed_fit}The distribution of \acp{SBH} for the
7.5~nm Au/30~nm Ag/Si(001)  (a) and  the 7.5~nm Au/1~nm Ag/Si(001)  (b) diodes
with multi-modal fits indicated in green.  Each normal distribution is shown in
black  and the mean \ac{SBH} of each distribution indicated by a dotted red
line. Inset in (a) is a schematic of the system modeled in the inset of (b),
which displays the relationship between patch size and calculated \ac{SBH}.  }

\end{figure}

The Schottky barrier height histograms for the silver samples can be utilized to
quantify the intermixing at the interface and compare to the metrology results.
The approach taken is to sum multiple Gaussians into one that fits the envelope
of the histogram and then compute the area under each Gaussian and assign it to
a species; Au, Ag, or defects.  It was found that five Gaussians were needed to
form a fitting envelope, two for each species and one for defects as seen in
Fig.~\ref{thick_histo_mixed_fit}.  The two for each species are attributed to
pinch-off effects due to the presence of different metals in close proximity to
one another~\cite{tung:mser:35:1, tung:prb:45:509, tung:jvstb:11:1546,
tung:prl:84:6078}. Periodically sampling over an array of patches as in this
study,  will result in several barrier heights in between that of pure Au
(0.84~eV) and pure Ag (0.66~eV)~\cite{balsano:aipa:3:112110}. Calculations
indicate that the change in the local barrier height between a circular patch of
Ag surrounded by Au transitions smoothly from the center of a silver patch to
the edge, which is supported with the experimental observation by Olbrich et al.
and Sirringhaus et al.~\cite{olbrich:jap:83:358, olbrich:apl:70:2559,
sirringhaus:prb:53:15944}. For an array of circular patches surrounded by and
infinite area of gold, numerical calculations indicate that the barrier height
inside and outside a circular silver patch is lower than for an isolated patch.
In addition, defects and/or foreign species will result in measuring higher
barrier heights such as those from oxides or simply from increased
scattering~\cite{sirringhaus:prb:53:15944}.

For the 30~nm Ag sample, the curves above 0.85~eV are attributed to both defects
and gold. The relative area under the fits to these peaks is $18.5\%$, putting
an upper bound on the presence of defects and gold. The area under the three
curves below 0.85~eV are attributed to Ag. Utilizing the perturbation equation
for a single circular \ac{SBH} inhomogeneity measured at its center, it is found
that these barrier heights correspond to isolated silver patches with radii of
200~nm and 400~nm, respectively as displayed in the insets of
Fig.~\ref{thick_histo_mixed_fit}~\cite{tung:mser:35:1, tung:prb:45:509,
tung:jvstb:11:1546, tung:prl:84:6078}. These sizes are larger than the sizes of
the regions in the maps, but are in agreement with the close proximity of the
patches.  The lowest peak is attributed to silver.  The  amount of silver at the
interface is approximated by the sum of the three lower peaks of $81.5\%$.

Applying similar analysis to the peaks of the 1~nm Ag sample indicates Au as the
dominant species at the interface by computing the area under the peaks above
0.85~eV. The peak around $0.85$~eV is considerably narrow indicating that the
contributions due to gold are in excess of $58\%$.  The large overlap between
the peak centered at 0.85~eV and 0.75~eV indicate that the latter peak has
contributions from both Au and Ag. The lowest two peaks are contributed to by
almost silver exclusively. These barrier heights are lower than expected from
the electrostatic model for a single patch, and are attributed to the close
proximity of the patches and a high level of intermixing of the species.  The
amount of silver at the interface is again the sum of the three lower peaks of
$26.5\%$.

These \ac{beem} estimates are consistent with the \ac{XPS} data, which shows
silver composition of about 90\% and 15\% at the interface for the 30~nm Ag and
1~nm~Ag samples, respectively. The \ac{beem} and XPS estimates are also
consistent with the large difference in the amount of silver between both
samples. The silver in 1~nm sample is most likely not a full layer and would
easily allow for Au at the interface.  The Au present at the interface in the
30~nm thick silver is attributed to diffusion and the miscibility of Au and Ag.

\section{Conclusion}

The \acl{SBH} inhomogeneities in mixed Au/Ag/Si(001) diodes has been mapped to
nanoscale dimensions using \ac{beem}. The average \ac{beem} spectra and fits
indicate that a single threshold model cannot capture the complex mixing
occurring at these interfaces.  These results show that \ac{SBH} maps and
histograms are needed to gain insight into these mixed interfaces, where fits to
the histograms are utilized to estimate the relative amount of species present
at the interface and are consistent with the chemical composition of the
interface observed with \ac{XPS} and \ac{RBS}.  These findings demonstrate a
method to infer interface composition from the interface electrostatic
properties as measured with \ac{beem}.

\section{Acknowledgments}

The authors acknowledge the support of the Semiconductor Research Corporation,
Center for advanced Interconnect Science and Technology, and the National
Science Foundation Grant DMR-1308102, and SEMATECH.

\pagebreak

\pagebreak

\pagebreak

\pagebreak

\pagebreak

\pagebreak

\end{document}